# Tailorable optical scattering properties of the V-shaped plasmonic nano-antennas: a computationally efficient and fast analysis

Arash Rashidi,[1] M. T. Chryssomallis,[2] and D. E. Anagnostou [3,*]

[1]*Electrical and Computer Engineering Department, University of Wisconsin-Madison, Madison, WI 53706, USA*
[2]*Electrical and Computer Engineering Department, Democritus University of Thrace, Xanthi, 67100, Greece*
[3]*Electrical and Computer Engineering Dept., South Dakota School of Mines & Technology, Rapid City, SD, 57701, USA*
**Corresponding author:* danagn@ieee.org*

We introduce an efficient computational scheme based on Macro Basis Function (MBF) method, to analyze the scattering of a plane wave by the V-shaped plasmonic optical nano-antennas. The polarization currents and the scattered fields for symmetric and anti-symmetric excitations are investigated. We investigate how the resonant frequency of the plasmonic V-shaped nanoantenna is tailored by engineering the geometrical parameters and by changing the polarization state of the incident plane wave. The computational model presented herein is faster by orders of magnitude than commercially available finite methods and is capable to characterize also other nano-antennas comprising of junctions and bends of nanorods.

## 1. INTRODUCTION

Plasmonic nanoantennas have recently found applications in photo-voltaic devices, bio-sensing, nonlinear optics, quantum optics and optical circuits [1,2]. Wire plasmonic nano-antennas which allow only one scattered electric field component along the nanorod axis have recently been studied and characterized [3,4]. However, controlling the polarization state of photons is required in some optical applications, e.g., cryptography, optical computing and communications [5]. Cross resonant plasmonic nano-antennas consisting of two perpendicular plasmonic dipole antennas have been shown to be capable to convert propagating fields of any polarization state into correspondingly polarized, localized and enhanced fields and vice versa [6,7]. Other plasmonic nanoantennas comprising of junctions that connect six or eight monopoles have been introduced to obtain a broadband spectral response when illuminated with circular and elliptical polarizations [8].

Recently there has been significant interest in V-shaped plasmonic nano-antennas because of their special phase response by spatially tailoring their geometry, i.e. L and Δ, in an array [9]. Moreover, arrays of plasmonic V-antennas have been shown to be capable of molding the wave-front of reflected and refracted beams in a nearly arbitrary way. Other advantages of the V-shaped plasmonic nanoantennas are the ease of fabricating planar antennas of nanoscale thickness and that such V-antennas consist of plasmonic nanorod resonators have widely tailorable optical properties [10,11].

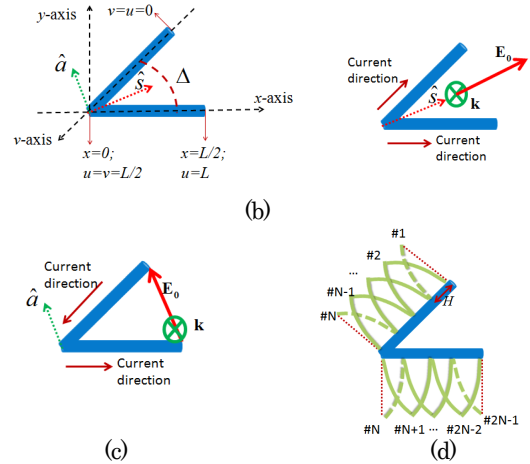

Fig. 1. (a) Geometry and coordinate system of V-shaped plasmonic nanoantennas, (b) Illustration of symmetric mode, (c) Antisymmetric mode, (d) Discretization of the polarization current.

To illustrate versatile polarization states for photons on plasmonic wire nano-antennas, we first consider a set of non-collinear monopoles. Each monopole radiates an electric field polarized in parallel to the corresponding axis. Because the monopoles are not collinear, different polarization states for the total scattered field are achieved. The simplest way to get a non-collinear set of coupled monopoles is introducing a V-shaped plasmonic nanoantenna consisting of two arms of equal length connected at one end at an angle Δ as illustrated in Fig. 1.

To investigate the importance of the V-shaped plasmonic antenna as a fundamental nano-element in plasmonic optics in changing the polarization state and phase of photons while giving rise to the geometrically tunable resonances to get a strong enough magnitude of the scattered field, we need to analyze efficiently and fast its optical performance. Also, to facilitate the design of optical metamaterials comprising of the V-shaped nanoantennas, an efficient formulation is needed to quantify their optical performance. Conventional EM-Simulation packages, which usually employ the Finite Difference Time Domain (FDTD) technique (or other algorithms in the time or frequency domain), require an expertise in using them as experimental results to validate the simulations can be difficult to obtain, while they are highly time- and memory-consuming, mainly because FDTD requires a fine meshing to characterize the thin cross-sections of the nanorod comprising of dispersive and negative permittivity $\varepsilon_r$ material [12,13].

The Macro Basis Function (MBF) method has proven to be a time- and memory-efficient technique for analyzing scattering from a plasmonic nanorod antenna [13] and also for the scattering analysis of elements comprising of junctions of wires and strips in microwave regime [14]. The main advantages of utilizing the MBF method are: 1) relatively small sized and well-conditioned matrix equation, 2) being faster in speed and requiring less memory, by orders of magnitude, compared to the conventional numerical methods, 3) it is a computationally efficient scheme to model the current at the junction of elements, e.g., cross-shaped elements, by satisfying Kirchhoff Current Law (KCL) and obviating the need to handle fictitious singularities of the electric field generated by piecewise sinusoidal Basis Functions (BFs) at the junction [14], 4) it demonstrates a close-form solution to the radiated electric field generated by MBFs [13], and 5) being compatible with the Characteristic Basis Function Method (CBFM) to efficiently characterize the scattering performance of infinite and large finite clusters of metamaterials [15]. In this letter we use the MBF method to analyze the scattering performance of the V-shaped plasmonic nanoantenna to take advantage of all the above mentioned features.

## 2. SCATTERING FROM A PLASMONIC V-SHAPED NANOANTENNA

This section demonstrates the formulation for the problem at hand. Fig. 1(a) demonstrates a V-shaped plasmonic nanoantenna located in $xy$-plane, which consists of two arms of equal length $L/2 = 75$ nm and equal radius $a = 7.5$ nm connected at one end at an angle $\Delta$. One arm is along $x$-axis and the other arm is along $y$-axis. We parameterize the combination of the $y$- and $x$-axes with the $u$ parameter that $0 \leq u \leq L$ (see Fig. 1(a)). Both arms are made of silver by which its permittivity is characterized by the Drude model: $\varepsilon_r = \varepsilon_{r\infty} - f_p^2 / [f(f - jf_d)]$, with $\varepsilon_{r\infty} = 5$, $f_p = 2175$ THz and $f_d = 4.35$ THz, where $f_p$ is the plasma frequency and $f_d$ the damping frequency [13]. We define two unit vectors to describe the orientation of a V-antenna: $\hat{\mathbf{s}}$ along the symmetry axis of the antenna (the bisector of the angle $\Delta$) and $\hat{\mathbf{a}}$ perpendicular to $\hat{\mathbf{s}}$ as in Fig. 1(a).

V-antennas support "symmetric" and "anti-symmetric" modes (Figs. 1(b,c)) [9]. Physically for a single plasmonic nanorod antenna illuminated by an arbitrarily incident plane wave [13], the longitudinal polarization current is excited by the vector component of the incident electric field which is directed along the axis of the nanorod (we can neglect the effect of the transverse polarization current in the cross section of the nanorod because of the thin cross section). Keeping this in mind, for the symmetric excitation, when the incident electric field is along $\hat{\mathbf{s}}$, (see Fig. 1(b)), the polarization current in each arm is directed from the end of the junction toward the open end of the corresponding arm. Similarly for the anti-symmetric excitation, when the incident electric field is along $\hat{\mathbf{a}}$ (Fig. 1(c)), by projecting the incident electric field onto each arm, the direction of the currents is along $\hat{x}$ (for the $x$-arm) and $\hat{y}$ (for the $y$-arm), respectively. The incident wave vector $\mathbf{k}$ for both symmetric and anti-symmetric modes is inward and normal to the plane of device, i.e., $-\hat{\mathbf{z}}$ (see Figs. 1(b,c)).

One way to solve the scattering performance of a dielectric body of relative permittivity $\varepsilon_r$ illuminated by the incident electric field $\mathbf{E^{Inc}}$, is by using the volume equivalent theorem [17]. In this theorem we replace the dielectric material with a polarization current density $\mathbf{J}$. The electric field radiated by such a polarization current density is then equivalent to the field $\mathbf{E^{Scat}}$ scattered by the original dielectric body. The polarization current density is proportional to the total electric field $\mathbf{E^{Tot}}$, the summation of the incident and scattered electric field, as below

$$\mathbf{J} = j\omega\varepsilon_0(\varepsilon_r - 1)\mathbf{E^{Tot}}, \quad (1)$$

where $e^{j\omega t}$ time dependency for the electric field and polarization current is assumed and suppressed. The first author has applied the volume equivalent theorem to efficiently characterize the optical scattering performance of plasmonic nanorods in [13]. We use the same theorem in this paper for scattering analysis of the V-shaped plasmonic nanoantennas.

By using (1), the equivalent polarization current on the V-antenna, flowing along $y-$ and $x-$ directions for the $y-$ and $x-$ arms, respectively, satisfies the following set of polarization equations [13]:

$$\begin{cases} E^{scat}_{yy} + E^{scat}_{yx} + E^{inc}_y = \zeta I_y(y) & \text{on } y\text{-arm} \\ E^{scat}_{xy} + E^{scat}_{xx} + E^{inc}_x = \zeta I_x(x) & \text{on } x\text{-arm} \end{cases}, \quad (2)$$

where $E^{Scat}_{wt}$ is the $w-$ component of the electric field scattered by the $t-$ arm, $t, w \in \{y, x\}$, which is measured along a line-segment on the top of the surface of the $t-$ arm parallel to its axis. Because the normal to the plane of the V-antenna is $z-$ axis (see Fig. 1), so the location of the observation line-segment is at $z = a$, i.e., the distance between the observation line-segment on the $t-$ arm and the axis of the $t-$ arm is equal to the radius of the nanorod. $E^{inc}_t$ is the $t-$ component of the incident electric field on the $t-$ arm. The $I_t(t)$ in (2) is the polarization current flowing on the axis of the $t-$ arm which is an unknown function yet to be determined. And finally, $\zeta = -j\eta/[\pi a^2 k(\varepsilon_r - 1)]$, where $\eta$ and $k$ are intrinsic impedance and the wave-number of the free space, respectively.

To solve (2) using the MBF method, according to the process presented in [13], we expand the polarization current in terms of $2N - 1$ piecewise sinusoidal MBFs (see Fig. 1(d)). The first and the last MBFs, #1 and #$2N - 1$, are half MBFs with the domain width of $H$ to model the nonzero current at the two ends of the V-antenna. Other MBFs are full triangular sinusoidal basis functions with the domain width of $2H$. The MBF#$N$ at the junction is the composite-MBF comprising of a half-MBF on the $y-$ arm and the other half-MBF on the $x-$ arm with equal current at the junction, i.e., $I_y(y = L/2) = I_x(x = 0)$, which automatically satisfies the KCL.

The details about the current distribution and the radiated electric field associated with the piecewise sinusoidal MBFs have been discussed in [13]. However, for the reader's convenience, we present the resultant formulas herein. Let us consider a full triangular-sinusoidal MBF including left and right half MBFs as shown in Fig. 2. In this example, the axis of the nanorod is along the $z$–axis, therefore for the problem at hand of the V-shaped nanoantenna, the proper coordinate rotations must be carried out to use the electric filed expressions that are presented, for the MBFs of the structure as shown in Fig. 2. Here, the current is assumed to exist only at the axis of the rod along the $z$–direction. The current is assumed to have a triangular-sinusoidal variation given by: $I_z = \sin[\beta(H-|z'|)]$, where $-H < z' < H$. The $z$–component of electric field radiated by the right half MBF, $0 < z' < H$, is given by:

$$G_{zz} = \frac{-j\eta}{4\pi}\left[\frac{e^{-jkR_1}}{R_1} - \frac{e^{-jkR_0}}{R_0}\left(\cos(kH) + j\frac{z}{R_0}\sin(kH)\right)\right.$$
$$\left. - z\sin(kH)\frac{e^{-jkR_0}}{kR_0^3}\right], \quad (3)$$

where $R_0$ and $R_1$ are the distances from the observation point to the center ($z'=0$) and the right end ($z'=H$) of the MBF depicted in Fig. 2. Similarly, as derived in [13], the $x$–component of the electric field radiated by the right half MBF depicted in Fig. 2 is as below

$$G_{xz} = \frac{j\eta x}{4\pi\rho^2}\left\{z\left[\frac{e^{-jkR_1}}{R_1} - \frac{e^{-jkR_0}}{R_0}\left(\cos(kH) + j\frac{z}{R_0}\sin(kH)\right)\right]\right.$$
$$\left. + \sin(kH)\rho^2\frac{e^{-jkR_0}}{kR_0^3} - H\frac{e^{-jkR_1}}{R_1}\right\}. \quad (4)$$

For the $y$–component of the radiated electric field by the right half MBF depicted in Fig. 2, it is sufficient to replace the observation component $y$ by $x$. The electric field radiated by the left half MBF depicted in Fig. 2 (for which $-H < z' < 0$) is obtained from (3-4) when $H$ is replaced by $-H$.

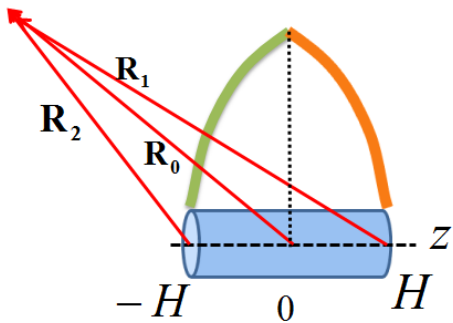

Fig. 2. Full-triangular sinusoidal MBF including right and left MBFs. The current flows along the axis.

After expanding the arm currents $I_v(v)$ and $I_x(x)$ in terms of MBFs defined in Fig. 1, we expand the electric fields radiated by the arm currents, $E_{wt}^{Scat}$ where $t, w \in \{v, x\}$, in terms of the electric field radiated by each MBF, by using the closed form formulas presented in (3-4) and the corresponding rotation of coordinates system. We therefore have $2N-1$ unknowns associated with the $2N-1$ weighting coefficients of MBFs. If we multiply both sides of (2) by each of the MBFs defined in Fig. 1, and take the integration along the MBF domain, Galerkin's testing [13], we achieve a $(2N-1) \times (2N-1)$ matrix equation. The integration in the Galerkin's testing is carried out by using the Gaussian Quadrature Rule (GQR) [13]. For each arm 20 observation points in GQR render the scheme convergent. Solving the resultant matrix equation leads to finding the polarization current and, therefore, we compute the scattered field by the nano-antenna. It is important to mention that the size of the matrix equation to reach to a convergent result is relatively small, for example only a 5×5 matrix equation (for 100 THz ≤ $f$ ≤ 400 THz) or a 7×7 matrix equation (for 400 THz < $f$ ≤ 600 THz) can efficiently and accurately model the scattering from the V-shaped plasmonic nano-antenna.

## 3. SYMMETRIC AND ANTISYMMETRIC EXCITATIONS

For the symmetric and anti-symmetric excitations at $f = 200$ THz and for different values of angle Δ, the magnitude and phase of the polarization currents are plotted versus $u/L$ in Figs. 3 and 4. In Fig. 3(a), the current is zero at the junction for the symmetric excitation. This happens because the KCL is set at the junction and also due to the symmetric excitation (see Fig. 1(b)). Only a zero current can be present at the junction because of these two constraints. The other observation in Fig. 3(a) is that for Δ=180°, in which the two arms are collinear and the V-antenna degenerates to a nanorod, there is no polarization current generated by the symmetric excitation. This is because, in this case, the incident electric field is perpendicular to the axis of the nanorod and is not coupled to any longitudinal plasmonic mode in the nanorod.

In Fig. 3(b), for the symmetric excitation, the phase of the polarization current jumps as much as 180° at the junction which is consistent with the physics, i.e., KCL and symmetric excitation, described above. On the other hand, the magnitude of the current for the anti-symmetric case, in Fig. 4(a), has a peak value at the junction and the phase at the junction is continuous and almost flat over the V-antenna for $0 \leq u \leq L$. The condition number for the 5×5 impedance matrix is only 35.8 at $f$ =200 THz for $\Delta$=30° which is a well-conditioned matrix. The condition number is even better for greater angles, e.g., for $\Delta$=180°, the condition number is only 6.8.

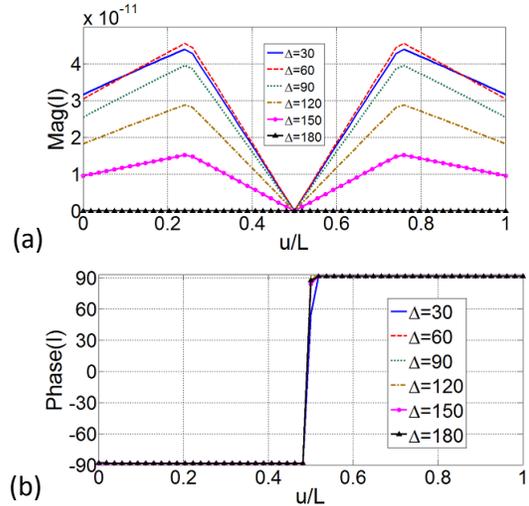

Fig. 3. Polarization current of the V-shaped plasmonic antenna illuminated by a symmetric excitation at $f$ =200 THz ($\lambda$ =1.5 μm) for different values of the angle Δ. (a) Magnitude, (b) Phase.

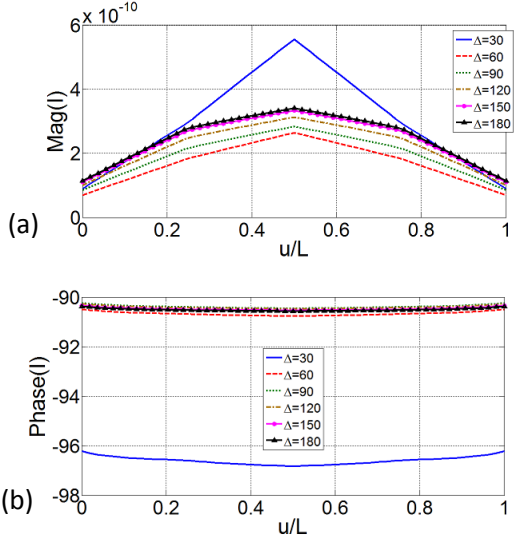

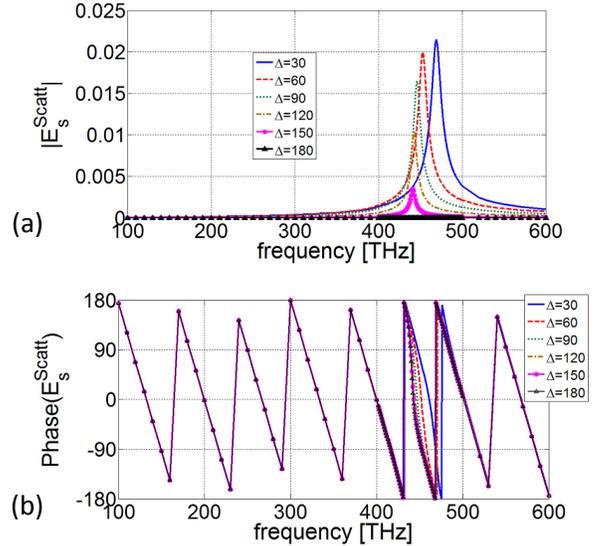

Fig. 4. Polarization current of the V-shaped plasmonic antenna illuminated by an anti-symmetric excitation at $f$ =200 THz ($\lambda$ =1.5 µm) for different values of the angle $\Delta$. (a) Magnitude, (b) Phase.

Fig. 5. Scattered field from the V-shaped plasmonic antenna illuminated by a symmetric excitation for different values of the angle $\Delta$. (a) Magnitude, (b) Phase.

The scattered field by the plasmonic V-antenna is the electric field radiated by the determined polarization current. Traditionally the electric field radiated by the given current can be found by performing the convolution of the free space Dyadic Green's Function (DGF) and the current. Our computational model obviates the need of using DGFs, because the polarization current in the V-antenna is a superposition of piecewise sinusoidal MBFs, and that the electric field radiated by each MBF is obtained in a closed form formula, as described before via equations (3-4).

Figs. 5 and 6 illustrate the magnitude and phase of the scattered field by the plasmonic V-shaped nanoantenna observed at $\mathbf{r} = 4.5[\mu m]\hat{\mathbf{z}}$ ($3\lambda$ at 200 THz) for symmetric and anti-symmetric modes, respectively. It is interesting to note that for the plasmonic V-antenna, the polarization state of the scattered photons is the same as that of the incident light when the latter is polarized along $\hat{\mathbf{a}}$ or $\hat{\mathbf{s}}$. This means the scattered field is symmetric when the V-antenna is illuminated by a symmetric excitation and is anti-symmetric when illuminated by an anti-symmetric incident field. This property of the plasmonic V-antenna allows one to design the polarization of the scattered light [9].

In Fig. 5(a), for the symmetric mode, the first-order resonant frequency for which the magnitude of the scattered field is maximum, occurs when the length of each arm is around one half of the effective wavelength, i.e., $L/2 = \lambda_{eff}/2$, because according to the Fig. 3, the current distribution in each arm approximates that of an individual straight antenna of length $L/2$ [9].

According to the approximated algebraic formula for the effective wavelength of a plasmonic nanorod given by Novotny [16], the first-order resonant frequency for the symmetric mode of the silver V-antenna studied herein is around $f_{res}$ =415 THz.

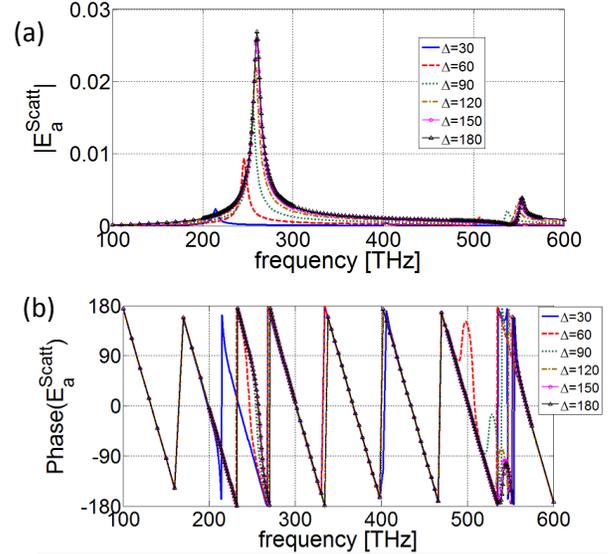

Fig. 6. Scattered field from the V-shaped plasmonic antenna illuminated by an anti-symmetric excitation for different values of the angle $\Delta$. (a) Magnitude, (b) Phase.

However, the Novotny algebraic equation is an approximate formula, and according to our computation depicted in Fig. 5(a), $f_{res}$ varies from 469 THz (for $\Delta$=30°) to 441 THz (for $\Delta$=150°). In Fig. 4(b), we observe that the phase of the scattered field across the resonance changes appreciably which makes the plasmonic V-shaped antenna capable to be utilized as an optical resonator to design the magnitude, phase and the polarization state of the scattered light [9]. Similarly, for the anti-symmetric excitation, one can conclude that the first-order resonance occurs when $L = \lambda_{eff}/2$ which is roughly $f_{res}$ =292 THz. More accurately, according to Fig. 6(a), for the anti-symmetric excitation, $f_{res}$ varies from 214 THz (for $\Delta$=30°) to 260 THz (for $\Delta$=180°). To facilitate the observation of appreciable phase variation of the scattered field around the resonance, in Fig. 7 for anti-symmetric case the phase of the scattered field shifted from that of the scattered field for $\Delta$=180°, i.e., $\angle E_a^{Scatt}(\Delta) - \angle E_a^{Scatt}(\Delta = 180°)$ is plotted versus the frequency for different $\Delta$ angles.

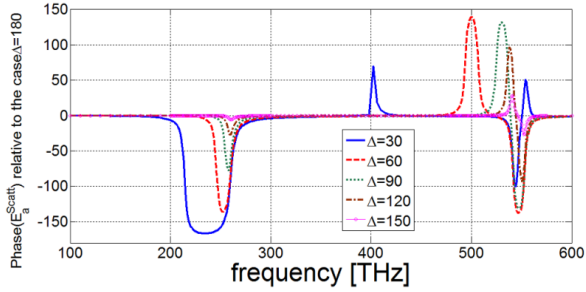

Fig. 7. The phase of the scattered field for anti-symmetric case shifted from the reference phase of the scattered field for $\Delta=180°$.

As a summary, by changing the polarization of the incident electric field from symmetric to anti-symmetric, the resonant frequency of the V-shaped antenna shifted significantly. Moreover, the resonant frequency can be tuned by changing the angle $\Delta$ or the length $L$, i.e., we get two degrees of freedom. This is the geometrically tunable double-resonance property of the V-shaped plasmonic nano-antenna.

The other interesting observation in Figs. 5 and 6 is that by decreasing the angle $\Delta$, the resonant frequency increases (blue-shift) for the symmetric mode, and decreases (red-shift) for the anti-symmetric mode. For the special case in which $\Delta$ is small, because of the current directions for the arms connected at the junction, for the symmetric mode decreasing the angle $\Delta$ is corresponded to have an equivalent nanorod with the higher effective radius because the currents are in the same direction (see Fig. 1(b)). This observation agrees with the result of the algebraic formula for the effective wavelength for a plasmonic nanorod given by Novotny [16], in which increasing the radius of the nanorod gives rise to a blue-shift. For the anti-symmetric mode with a small $\Delta$ by considering the opposite directions for the current at the junction (Fig. 1(c)) the composite-arm will have the smaller equivalent radius (because the joint currents cancel each other) which gives rise to a red-shift. Also because the currents tend to cancel each other (destructive effect) the peak of the scattered field for the small $\Delta$ in the anti-symmetric mode decreases significantly (see Fig. 6(a)).

## 4. SCATTERING FROM PLASMONIC V-ANTENNA

Next we compare the results with those obtained from the commercial software CST Microwave Studio (MWS) using both time and frequency domain solvers. The plasmonic V-antenna is made of silver whose permittivity is characterized by the Drude model that described in section 2. The arm length of the V-antenna is $L/2 =75$ nm and the radius is $a =7.5$ nm. The angle of V-antenna is equal to $\Delta=60°$. The excitation is plane-wave with the symmetric polarization (as demonstrated in Fig. 1(b)). The frequency range for the incident plane wave is $100\,\text{THz} < f < 600\,\text{THz}$. The total (the sum of the scattered and incident) field is measured by defining the electric fields probes in CST MWS at $\mathbf{r} = 0.2[\mu m]\hat{\mathbf{z}}$ which is $\lambda/2.5$ at higher frequency $f =600$ THz which is in the near field region for the frequency range mentioned above. The boundary condition is open for all the six faces of the boundary box. The faces of the boundary box are $\lambda/4$ (at the lower frequency $f =100$ THz) away from the origin (the junction of the V-antenna).

For the time domain solver in CST MWS, we have used ~28 million hexahedral mesh-cells. In the mesh properties part of the CST MWS, we set 'Line per wavelength' equal to 80 instead of the default value, which is 10. Moreover, in the CST local mesh properties, we define the local edge refinement factor equal to 10, instead of the default value 1, just to make sure that the thin arms of the plasmonic V-antenna are discretized very well in order to reach to a computational convergence for the energy convergence criterion equal to -80 dB level. The CPU running time for the CST time domain solver is around 2 days on an Intel(R) Xeon(R) CPU having two 2.40 GHz processors. However it takes only 1 sec/200 frequency samples for the MBF method on the same machine to compute the scattered field on the entire frequency band regardless of the location of the observation point.

For the frequency domain solver in CST MWS, we have used ~800,000 tetrahedral mesh-cells. In the mesh properties part of the CST MWS, we set 'steps per wavelength' equal to 6 instead of the default value, which is 4., just to make sure that the thin arms of the plasmonic V-antenna are discretized very well in order to reach to a computational convergence. The CPU running time for the CST frequency domain solver is around 1 day on the computer described above.

Fig. 8 demonstrates the comparison between the scattered field obtained by MBFM, CST time domain and CST frequency domain solvers. As we observe, the result of CST frequency domain solver is pretty matched to that of MBFM. For the CST time domain solver the only difference is the magnitude of the peak of the resonance. This difference is due to the fact that CST time domain solver needs a very fine mesh size for discretization of the arm cross section of the plasmonic-V-antenna. We already have the local mesh cell size equal to $a/24$ to discretize the cross section of the plasmonic arms. Making the mesh size finer will drastically increase the CPU running time.

The radiation patterns for the V-shaped antenna in the $x-y$ plane and in the plane including $\phi =120°$ and $\phi =300°$ are shown in Fig. 9 at $f = 450$ THz (resonance), for the symmetric excitation and for $\Delta=60°$. A good agreement between CST and MBF Method (MBFM) is observed. In Fig. 9, we observe that the radiation pattern of the V-shaped antenna illuminated by the symmetric plane-wave excitation has a donut shape whose symmetric axis is along the bisector of the V-antenna, i.e., $\phi =30°$. The physical reason is that because of the symmetric excitation, the current distribution on each arm is the same (in magnitude and phase) and is close to the current distribution of a resonant dipole antenna. Thus, the $x-$arm and the $v-$arm radiate donut-shaped patterns with the main lobes located at $\phi_1 =\{90°, 270°\}$ and $\phi_2 =\{150°,330°\}$, respectively in the $x-y$ plane. Hence, the resultant super-imposed pattern is a donut-shaped pattern with the main lobes placed at $\phi_{net} = (\phi_1 + \phi_2)/2 =\{120°, 300°\}$, according to Fig. 9(a) in the $x-y$ plane. The radiation pattern is omni-directional (see Fig. 9(b)) at the plane $\phi_{net}$ comprising of $\phi =120°$ and $\phi =300°$.

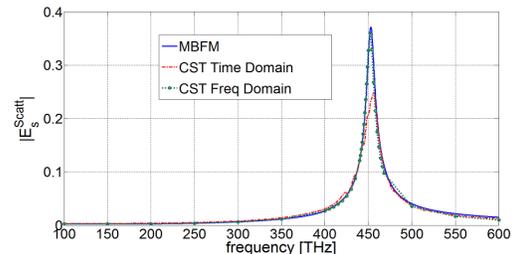

Fig. 8. Comparison of the MBF method and CST time and frequency domain solvers to compute the scattered field for the symmetric mode.

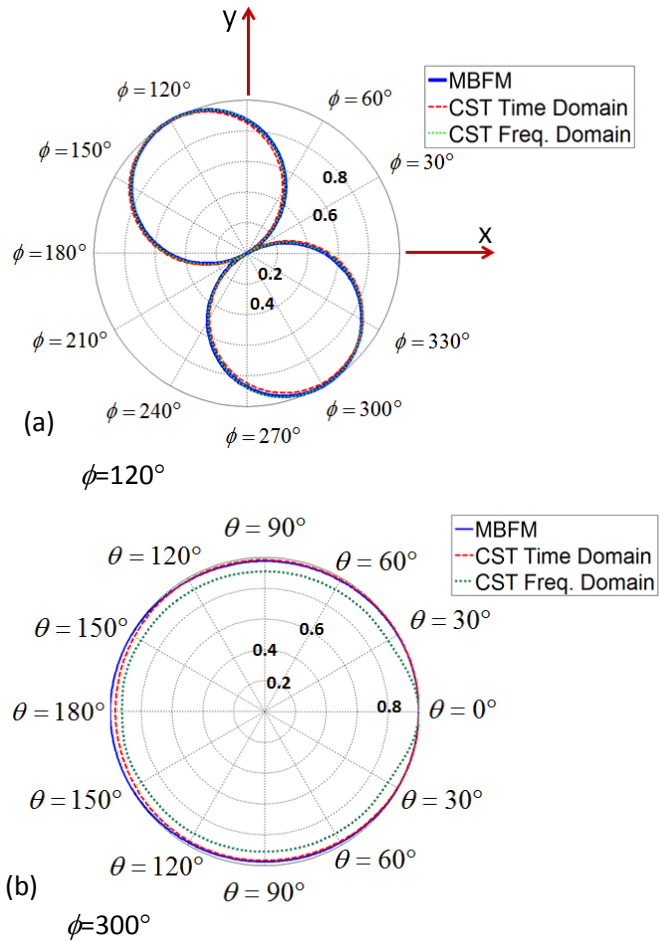

Fig. 9: Radiation pattern for the symmetric excitation and Δ=60° computed with MBF Method compared to CST. (a) At θ=90° (xy-plane), (b) At $\phi$ = 120° (upper-half-plane) and $\phi$= 300° (lower half-plane).

## CONCLUSIONS

In this paper we have introduced a computationally efficient, strong and fast technique to analyze the scattering of an optical plane-wave from a plasmonic V-shaped nano-antenna. Our technique is based on MBFM which leads to small-sized and well-conditioned matrix equation constructed by using closed form formula for the electric field radiated by piece-wise sinusoidal MBFs. The technique explained in our paper is orders of magnitude faster than the time domain and frequency domain solvers in CST MWS. Our computational scheme can be used as a powerful engine for efficient analysis and design optimization of large arrays of plasmonic configurations comprising of junctions of monopoles.

## ACKNOWLEDGEMENT

This work was supported by the Defense Advanced Research Projects Agency (DARPA)/Microsystems Technology Office (MTO) Young Faculty Award program under Award No. N66001-11-1-4145, the National Science Foundation under grants ECS-1310400, ECS-1310257, and IIP-1417284, and the Hellenic Ministry of Education project Thales RF-Eigen-Sdr. The authors thank Dr. M. Al-Tarifi for his comments and help with CST simulations.